# A Novel Strange Attractor with a Stretched Loop

Safieddine Bouali

University of Tunis, Management Institute,
Department of Quantitative Methods & Economics,
41, rue de la Liberté, 2000, Le Bardo, Tunisia
Safieddine.Bouali@isg.rnu.tn

*The paper introduces a new 3D strange attractor topologically different from any other known chaotic attractors. The intentionally constructed model of three autonomous first-order differential equations derives from the coupling-induced complexity of the well-known Lotka-Volterra oscillator. The chaotic attractor exhibiting a double scroll bridged by a loop mutates to a single scroll with a very stretched loop by the variation of one parameter. Analysis of the global behavior of the new low dimensional dissipative dynamical model and its range of periodic and a-periodic oscillations are presented.*

## 1. Introduction

In a seminal paper, Lorenz [1] found the first chaotic attractor in a three dimensional autonomous system while studying atmospheric instability. Its relatively simple nonlinear structure displays both sophisticated and stable dynamical behaviors.

Followed by Rössler [2], Chua [3, 4], Sprott [5], etc., chaos from nonlinear approach has been shown to be useful to overlap classical results in a global framework where strange chaotic patterns are possible conjectures. Yet, chaotic behavior probably exists and deserves a deep exploration. Indeed, in a wide range of disciplines, discovering chaos in low dimensional dissipative dynamical systems spreads research projects to redefine models from linear to nonlinear specifications.

Purposefully creating chaos can be a nontrivial task to focus a new kind of dynamical patterns. To this end, process of chaotification or anti-control of chaos [6, 7, 8, for example] explores the induced chaotic behaviors of an originally non-chaotic system via several applications.

We applied previously this derivation methodology of the Van der Pol 2D oscillator leading to a specific class of 3D strange attractors [9] implemented also in an electronic circuit [10].

Coupling-induced complexity focuses the founding, at best of our knowledge, another canonical class of chaotic systems, and no having a topologically equivalent structure, is the scope of the present research.

## 2. The Modelization

Methodology of feedback anti-control, unraveling chaos has been widely applied. It is the ''bridge'' to generate complexity triggering a mechanism of new dynamical patterns from elementary 2D models.
Indeed, we retain to chaotify the relatively simple, and well-established Lotka-Volterra 2D system, modelizing the interspecific competition between two species [11, 12, 13].

### 2.1. A Lotka-Volterra Like System

Suspending the peculiarity of the 2D Lotka-Volterra system, we find only its dynamical behavior expecting chaotic pattern through the linkage of an endogenous perturbation into a third dimension.
Firstly, a quadratic term of x is introduced in the y-equation of the Lotka-Volterra oscillator to extend its domain of variation:

$$\begin{cases} \dot{x} = x\,(a - y) \\ \dot{y} = -\,y\,(b - x^2) \end{cases}$$

where $x \in \mathbb{R}$ and $y \in \mathbb{R}^+$ are the state variables of the model.
This Lotka-Volterra-Like system (LVLS) leads to a two stable and symmetric limit-cycles depending strongly from the initial conditions, where $(x_0, y_0) \neq (\pm\sqrt{b}, a)$.
The extension to the third dimension of the LVLS far from its "natural" oscillation could be obtained by injecting perturbation through an anti-equilibrium z-equation stressing the "robustness" of the system. One would expect the emergence of complexity and chaotic patterns in our application from this connection.

### 2.2. A New Chaotic System

The exploration of new dynamical behaviors leads us to connect a z-equation to the LVLS. Its arbitrary formulation, chosen between several specifications, is controlled by the parameter α.
We obtain a new system described by the following three-nonlinear differential equations:

$$\begin{cases} \dot{x} = x\,(4 - y) + \alpha\,z & \quad [\mathrm{I}] \\ \dot{y} = -\,y\,(1 - x^2) & \quad [\mathrm{II}] \\ \dot{z} = -\,x\,(1.5 - s\,z) - 0.05\,z & \quad [\mathrm{III}] \end{cases}$$

where x, y, and z, are the state variables of the system.
The model embeds two quadratic nonlinearities and only one cubic term, respectively, xy, xz, and yx². Requirement of simplicity leads us to select the minimum of parameters, i.e. s and α, two positive scalars.

For $P_0$ (α, s) = (0.3, 1), the obtained phase portrait displays a chaotic attractor with double scroll connected by a singular loop (Fig.1).

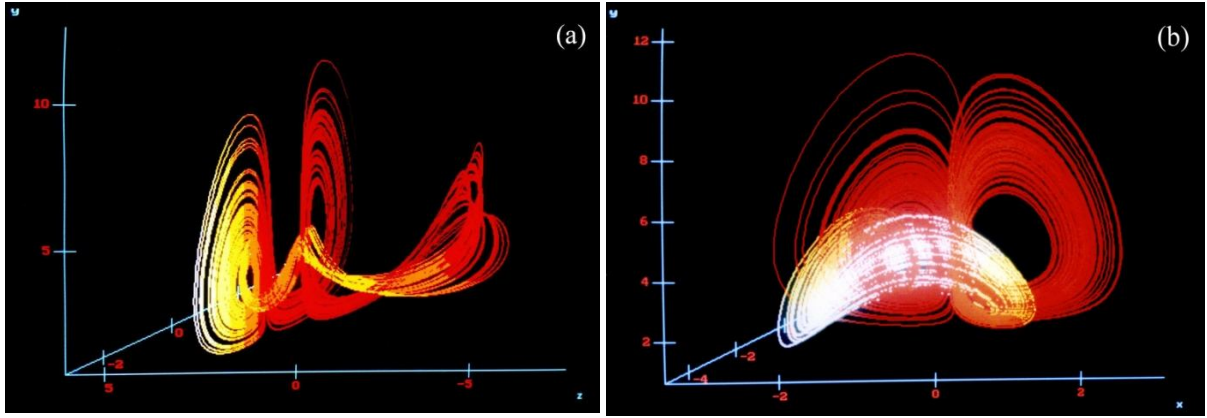

**Fig.1.** Phase portraits of the new chaotic attractor exhibiting its left and right scrolls bridged by a stretched loop for $P_0$ (α, s) = (0.3, 1). (a) left and (b) front images.

Changing only the parameter s, the attractor becomes different exhibiting a single wing with a very stretched loop (Fig. 2).

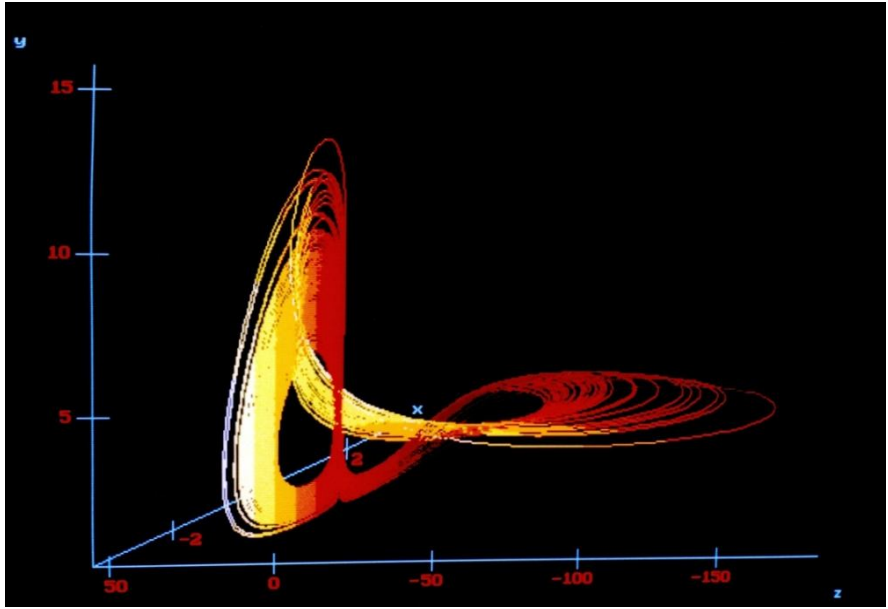

**Fig. 2.** An evolved single-left scroll attractor exhibiting an orthogonally very stretched loop using $P_1$ (α, s) = (0.3, 6).

The equilibrium points of these two attractors are found by setting $\dot{x} = \dot{y} = \dot{z} = 0$. Starting firstly from the set of parameters, $P_0$ (α, s) = (0.3, 1) and consecutively $P_1$ (0.3, 6), we obtain by solving [II], y= 0 or x = ± 1. While y= 0, its substitution in [I] gives $E_0$ ($x_0$, $y_0$, $z_0$) and $E_1$ ($x_1$, $y_1$, $z_1$). When x= ± 1, the two other equilibrium are namely $E_2$ ($x_2$, $y_2$, $z_2$) and $E_3$ ($x_3$, $y_3$, $z_3$).

The elementary attributes of these equilibria given by the corresponding eigenvalues $\lambda_i$ are found by solving the characteristic equation $|J - \lambda I| = 0$, where $I$, the unit matrix and $J$, the Jacobian of the model:

$$|J| = \begin{bmatrix} 4-y & -x & \alpha \\ 2xy & -(1-x^2) & 0 \\ sz-1.5 & 0 & sx-0.05 \end{bmatrix}$$

Coordinates, eigenvalues and features of stability related to the fixed points are reported in Table 1 and Table 2, respectively for $P_0 (\alpha, s) = (0.3, 1)$ and $P_1 (0.3, 6)$.

| Coordinates of the Equilibia | The corresponding characteristic equation, $|J - \lambda I|= 0$, and eigenvalues | Index [1] and Stability |
|---|---|---|
| $E_0 (x_0, y_0, z_0) = (0, 0, 0)$ | $\lambda^3 - 2.95 \lambda^2 - 5.50 \lambda - 1.55 = 0$<br><br>$\lambda_1 \approx 4.270$<br>$\lambda_2 \approx -1.028$<br>$\lambda_3 \approx -0.341$ | **Index-1**<br><br>Unstable:<br>*Saddle point* |
| $E_1 (x_1, y_1, z_1) = (0.062, 0, 0.833)$ | $\lambda^3 - 3.01 \lambda^2 - 5.98 \lambda - 1.90 = 0$<br><br>$\lambda_1 \approx 4.428$<br>$\lambda_2 \approx -1$<br>$\lambda_3 \approx -0.428$ | **Index-1**<br><br>Unstable:<br>*Saddle point* |
| $E_2 (x_2, y_2, z_2) = (1, 4.473, 1.578)$ | $\lambda^3 - 0.47 \lambda^2 + 8.52 \lambda - 8.49 = 0$<br><br>$\lambda_1 \approx 0.931$<br>$\lambda_2 \approx -0.265 + 2.99\,i$<br>$\lambda_3 \approx -0.265 - 2.99\,i$ | **Index-1**<br><br>Unstable:<br>*Spiral saddle point* |
| $E_3 (x_3, y_3, z_3) = (-1, 3.571, 1.428)$ | $\lambda^3 + 0.621 \lambda^2 + 7.11 \lambda + 7.49 = 0$<br><br>$\lambda_1 \approx -1$<br>$\lambda_2 \approx 0.19 + 2.730\,i$<br>$\lambda_3 \approx 0.19 - 2.730\,i$ | **Index-2**<br><br>Unstable:<br>*Spiral saddle point* |

(1) Index reports the number of eigenvalues with real parts Re $(\lambda) > 0$. From 1 to 3, it indicates the degree of instability. Index-0: null or negative real parts of all eigenvalues of the equilibrium characterize its stability.

**Table 1.** Equilibria of the system for $P_1 (0.3, 1)$

| Coordinates of the Equilibia | The corresponding characteristic equation, $\|J - \lambda I\|= 0$, and eigenvalues | Index [1] and Stability |
|---|---|---|
| $E_0\ (x_0, y_0, z_0) = (0, 0, 0)$ | $\lambda^3 - 2.95\ \lambda^2 - 3.75\ \lambda + 0.45 = 0$<br><br>$\lambda_1 \approx 3.888$<br>$\lambda_2 \approx -1.037$<br>$\lambda_3 \approx 0.099$ | **Index-2**<br><br>**Unstable:**<br>*Saddle point* |
| $E_1\ (x_1, y_1, z_1)= (0.010, 0, 0.138)$ | $\lambda^3 - 2.99\ \lambda^2 + 4.16\ \lambda + 0.20 = 0$<br><br>$\lambda_1 \approx -0.046$<br>$\lambda_2 \approx 1.473 + 1.458\ i$<br>$\lambda_3 \approx 1.473 - 1.458\ i$ | **Index-2**<br><br>**Unstable:**<br>*Spiral saddle point* |
| $E_2\ (x_2, y_2, z_2)= (1, 4.075, 0.252)$ | $\lambda^3 + 5.2\ \lambda^2 - 3.69\ \lambda - 48.49 = 0$<br><br>$\lambda_1 \approx 2.716$<br>$\lambda_2 \approx -3.958 + 1.465\ i$<br>$\lambda_3 \approx -3.958 - 1.465\ i$ | **Index-1**<br><br>**Unstable:**<br>*Spiral saddle point* |
| $E_3\ (x_3, y_3, z_3)= (-1, 3.925, 0.248)$ | $\lambda^3 - 5.97\ \lambda^2 - 8.30\ \lambda - 47.48 = 0$<br><br>$\lambda_1 \approx 2.849$<br>$\lambda_2 \approx -5.957$<br>$\lambda_3 \approx -2.791$ | **Index-1**<br><br>**Unstable:**<br>*Saddle point* |

(1) Index reports the number of eigenvalues with real parts Re $(\lambda) > 0$. From 1 to 3, it indicates the degree of instability. Index-0: null or negative real parts of all eigenvalues of the equilibrium characterize its stability.

**Table 2.** Equilibria of the system for $P_1$ (0.3, 6)

## 3. Global Behavior

In order to detect the dynamical behaviors of the system, a Diagram of Bifurcation is computed by varying the parameter s, considered as the "controller" of the system (Fig. 3). One can observe different dynamical patterns, within chaotic bubbles and the several windows of stability.

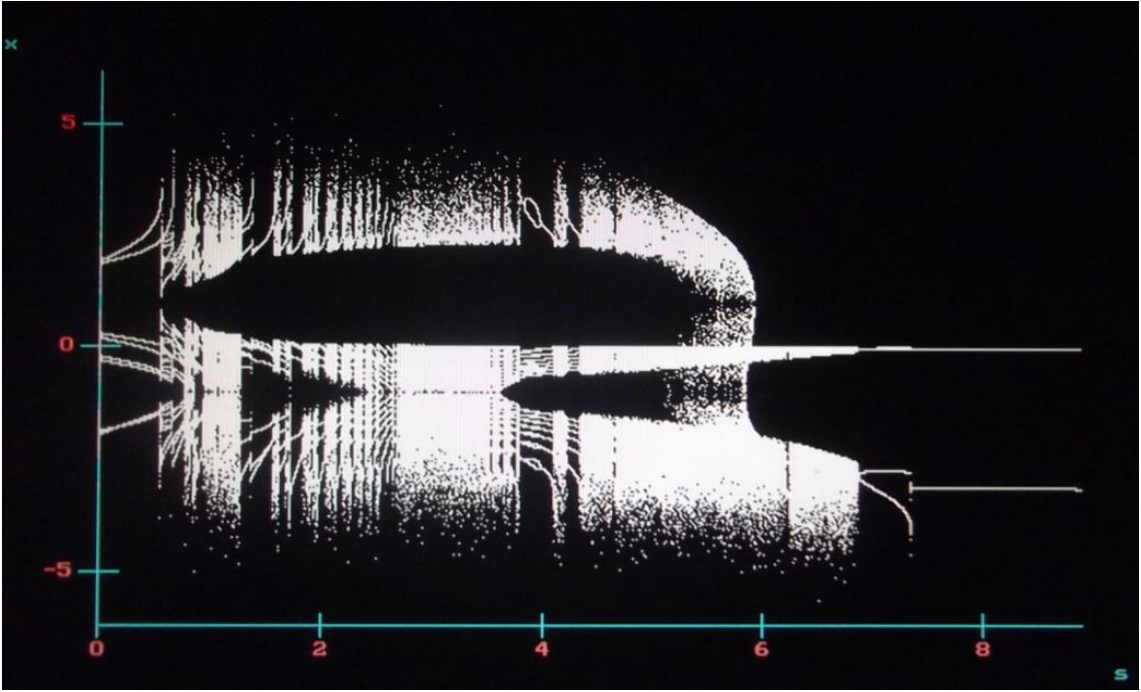

**Fig. 3.** Diagram of bifurcation of x for the trajectories cutting the plane (x, z) at the level $\bar{y} = 5$. Dots are related to both scroll and loop. Beyond s = 1, the attractor loses its second scroll but keeps its loop.

For example at s = 0.4, periodic orbit is observed (Fig. 4). We notice its shape very close to the chaotic attractor shown in fig.1. However, for the value s = 5.8, the shape of the displayed limit-cycle (Fig.5) is related to the chaotic attractor shown in fig. 2.

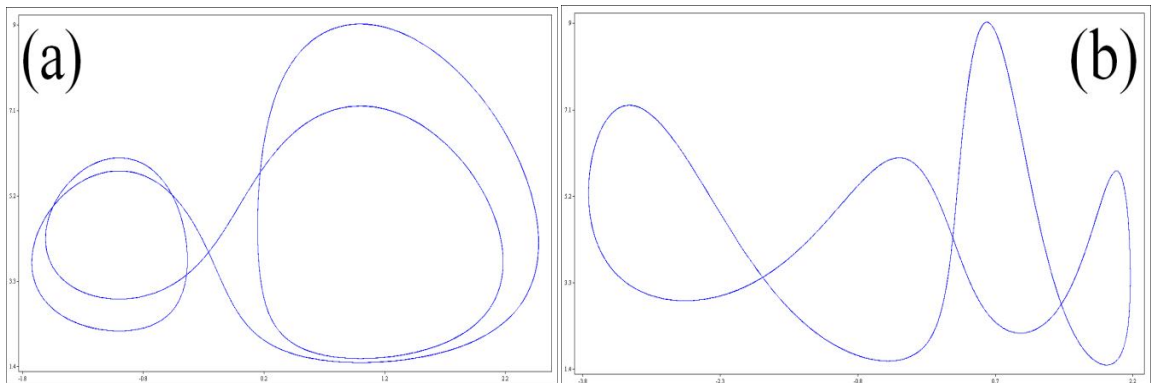

**Fig.4.** Projections of the limit-cycle for P (α, s) = (0.3, 0.4): (a) x–y phase plane, (b) y–z phase plane.

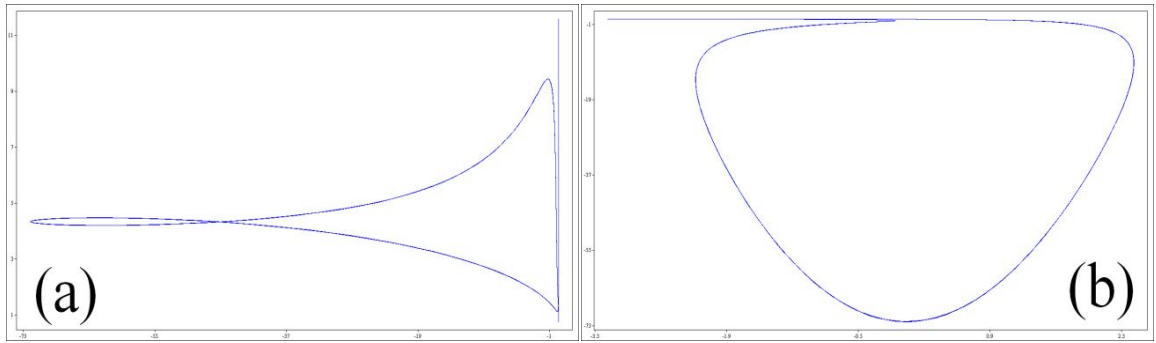

**Fig.5.** Projections of the limit-cycle for P (α, s) = (0.3, 5.8): (a) x–y phase plane, (b) y–z phase plane.

Eventually, does the new system have explicitly and globally a chaotic nature? It is known that the spectrum of Lyapunov exponents is the most useful diagnostic to quantify chaos. When the nearby trajectories in the phase space diverge at exponential rates, giving a positive Lyapunov exponent, the dynamics become unpredictable. Any system containing at least one positive Lyapunov exponent is defined to be chaotic. As can be seen from the Lyapunov spectrum for a varying parameter s in the range [0.2, 6.2], the new system is chaotic with a positive dominant exponent (Fig.6).

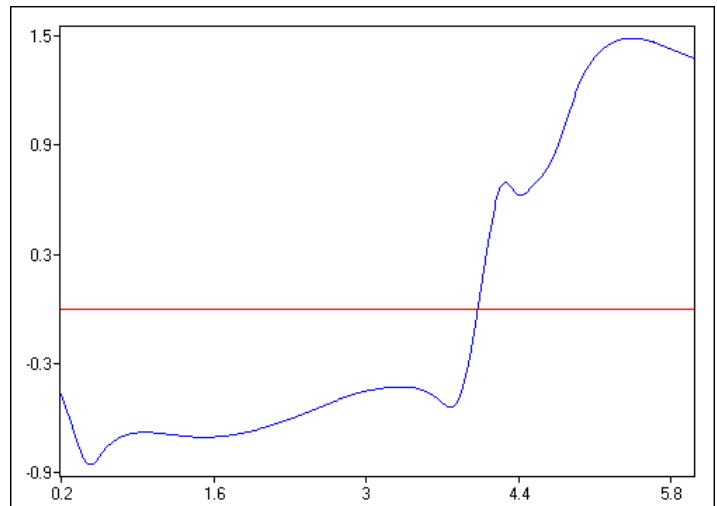

**Fig.6.** Largest Lyapunov Exponent for P (0.3, s) when s is varying in the range [0.2, 6.2].

## 4. Concluding Remarks

Several specifications of the feedback z-equation have been experimented. The simplest one, inducing the widest range of dynamical behaviors, has been selected.

The new chaotic system depicts a complex 2-scroll butterfly-shaped attractor exhibiting a singular loop. It mutates to a single wing, with a very stretched orthogonally loop, according to its sensitive dependency on parameters.

This intentionally constructed chaotic system doesn't reincarnate known strange attractors. The appearance of the new chaotic attractors is utterly different from the other existing chaotic systems showing double or a single scroll. Indeed, at the best of our knowledge, it can be verified that there does not exist diffeomorphism between the new system and the others chaotic 3D models (Lorenz-like systems, Rössler, Chua, Sprott…) since the eigenvalues structures of their corresponding equilibrium points are not equivalent.

Further developments of the analysis with an extended set of parameters will be investigated in a future work. In fact, the new system could be also suitable for digital signal encryption in the communication field when its variants provide a very large set of encryption keys. Reaching chaos through a relatively simple algebraic structure by chaotifying an originally stable oscillator, in the present case the LVLS, is still a challenging task.

*March 30, 2012*